\documentstyle[12pt, amssymb, epsfig,graphicx
]{article}

\def\fun#1#2{\lower3.6pt\vbox{\baselineskip0pt\lineskip.9pt
        \ialign{$\mathsurround=0pt#1\hfill##\hfil$\crcr#2\crcr\sim\crcr}}}

\renewcommand\[{\left[}

\newcommand\ee{\end{equation}}
\newcommand\be{\begin{equation}}
\newcommand\eea{\end{eqnarray}}
\newcommand\bea{\begin{eqnarray}}

\def\dslash{\not{\hbox{\kern-2pt $\partial$}}}
\def\Dslash{\not{\hbox{\kern-4pt $D$}}}
\def\Oslash{\not{\hbox{\kern-4pt $O$}}}
\def\Qslash{\not{\hbox{\kern-4pt $Q$}}}
\def\pslash{\not{\hbox{\kern-2.3pt $p$}}}
\def\kslash{\not{\hbox{\kern-2.3pt $k$}}}
\def\qslash{\not{\hbox{\kern-2.3pt $q$}}}

 \newtoks\slashfraction
 \slashfraction={.13}
 \def\slash#1{\setbox0\hbox{$ #1 $}
 \setbox0\hbox to \the\slashfraction\wd0{\hss \box0}/\box0 }

\def\ee{\end{equation}}
\def\be{\begin{equation}}

\renewcommand{\[}{\left[}

\newcommand{\beq}{\begin{equation}}

\newcommand{\eeq}{\end{equation}}
\newcommand{\beqa}{\begin{eqnarray}}
\newcommand{\eeqa}{\end{eqnarray}}

\def\simlt{\lesssim}
\def\simgt{\gtrsim}

\newcommand\slabel[1]{\label{#1}}

\newcommand\eq[1]{Eq.~(\ref{#1})}
\newcommand\eqs[2]{Eqs.~(\ref{#1}) and (\ref{#2})}
\newcommand\eqss[3]{Eqs.~(\ref{#1}), (\ref{#2}) and (\ref{#3})}
\newcommand\eqsss[4]{Eqs.~(\ref{#1}), (\ref{#2}), (\ref{#3})
and (\ref{#4})}

\begin{document}
\thispagestyle{empty}
\vspace*{.5cm}
\noindent
\vspace*{1cm}
\begin{center}
{\Large\bf The Amplitude of Dark Energy Perturbations
}\\[.8cm] {\large Christopher Gordon$^1$ and 
 David Wands$^2$}\\[.6cm]
{\it $^1$ Kavli Institute for Cosmological Physics, Enrico Fermi
  Institute and  Department of Astronomy and Astrophysics,
University of Chicago, Chicago IL 60637 \\
$^2$ Institute of Cosmology and Gravitation, University of Portsmouth,
Portsmouth PO1 2EG, United Kingdom}
\end{center}
\vskip 1 cm

\begin{abstract}
We propose a model which produces dark energy perturbations large
enough to explain the lack of power seen at the quadrupole scale in the
cosmic microwave background. If the dark energy is frozen from
horizon exit during inflation until dark energy domination, then it is
not possible to  
have perturbations in the dark energy which are large enough. We
propose using a tachyonic amplification mechanism to 
overcome this. The dark energy is
taken to be a complex scalar field, where the radial field has a
Mexican hat potential. During inflation, the radial component is
trapped near the maximum of its potential. At the end of inflation, it
rolls down to the minimum. The dark energy today is taken to be a 
pseudo-Nambu-Goldstone boson. The perturbations
generated during inflation are amplified by the rolling of the radial
field. We also examine the use of the variable decay mechanism in
order to generate an anti-correlation  between the dark energy
perturbations and the curvature perturbation. We show that using this
mechanism then constrains the properties of the dark energy and its
evolution from redshift one until today.
\end{abstract}



%

\section{Introduction}

Recently, it has been shown, that there is some evidence, that the dark
energy may not be constant on large spatial scales
\cite{MorTak03,GorHu04}. 
This is because a smooth dark energy on large
scales predicts a temperature/temperature (TT) quadrupole moment in the CMB
which is much larger than the observed WMAP value \cite{Ben03,Spe03}. The
original estimate was that the 
probability of having such a low or lower quadrupole given a
cosmological constant was only 0.7 $\%$ \cite{Spe03}. But, other
studies claim the probability is closer to $4\%$ \cite{lowquad}. 

In reference \cite{KawMorTak01} it was shown that if the dark energy has
perturbations which are initially uncorrelated with the perturbations in the
other matter components, then this will make the
(TT) quadrupole even higher, and so is not favored by
the data. Then, in reference \cite{MorTak03} it was shown that adding
perturbations which are spatially negatively correlated with the
initial perturbations in the other components can lower the estimated
TT quadrupole. In reference \cite{GorHu04}, the relative magnitude of the dark
energy perturbations $(\delta \rho_Q / \rho_Q)$ relative to the
curvature perturbation ($\zeta$), on constant density, or equivalently
comoving,  
hyper-surfaces 
\cite{Bar80,BarSteTur83,sepuni}, was parameterized as 
\beq
S \equiv {1\over \zeta} {\delta \rho_Q \over \rho_Q}.
\slabel{Sdef}
\eeq
where the values of the quantities are taken some time before
the onset of dark energy
domination, which occurs at $z\approx 1$. This parameterization includes the
possibility of no perturbations in the dark energy, which occurs when
$S=0$. 
It was found that 
\beq
{\rm Pr}(S > 0| {\rm TT}) = 0.04, \quad {\rm Pr}(S > 0|{\rm  TT,TE}) = 0.06
\eeq
where ${\rm Pr}(x|y)$ stands for the probability of $x$ given $y$ and
TT and TE are the WMAP temperature and temperature/polarization power
spectra. 
 The mean with
68\% confidence interval, using the TT and TE, data was 
\beq
S = -11.8 \pm 7.1.
\slabel{Smeas}
\eeq
The temperature data is already cosmic variance limited at the
quadrupole scale, but the polarization data is not, and future
measurements may more than double the amount of information available
 \cite{DorHolLoe04,GorHu04}.

As first shown in reference \cite{MorTak03}, for a specific model, and in
reference \cite{GorHu04}, under more general conditions, such large
perturbations in the dark energy density, 
as in \eq{Smeas},
imply a gravitational wave signal which is much larger than allowed by
observations. 
In this paper we propose a mechanism of tachyonically amplifying the
initial dark energy perturbations so as to reduce the gravitational
wave signal to negligible levels. 

In Sec.~\ref{background}, we review
in more detail why the gravitational waves are a problem. The basic
amplification mechanism is outlined in Sec.~\ref{amplification}. The
role of the inflaton field and its relation to the amplification
mechanism 
is investigated in Sec.~\ref{onephase} and Sec.~\ref{sepinflaton}. 
A variable decay reheating mechanism is discussed in Sec.~\ref{vardecay}.
Then,
in Sec.~\ref{example} an 
 example is given. The conclusions are summarized in
Sec.~\ref{Conclusions}.

\section{The problem of gravitational waves}
\slabel{background}
The continuity equation for a matter
component with equation of state $w_Q$, can be written as 
\beq
{{ \rho_Q'} \over \rho_Q} = 3 (1+w_Q)
\slabel{cont}
\eeq
where the prime indicates differentiation with respect to the number
of efolds of expansion which is given by
\beq
N \equiv -\log(a).
\eeq
If we set $a=1$ today, then at red shift one, when dark energy began to
dominate, we have $N \approx 0.7$. In general we will express time in
efolds as this will allow an easy conversion of background equations
into perturbation equations via the separate Universe approach
\cite{sepuni}. This method was originally used in estimating 
the curvature perturbation originating from inflation \cite{sepuni1}. 
However, it can also be used for modeling large scale perturbations at
 any cosmological epoch \cite{sepuni}. 

Current observational limits of the equation of state of dark energy
give \cite{Sel04}
\beq
w_Q \approx -1 \pm 0.1
\slabel{wQ}
\eeq
at the 68 $\%$ confidence interval 
 and so \eq{cont} implies that $\rho_Q$ is close
to constant from redshift one till today. Perturbing \eq{cont} gives
the large scale perturbation equation in the flat gauge
\cite{sepuni}. In this paper all perturbations of matter variables
will be in the flat gauge.
If we assume the equation of state of the dark energy
is unperturbed then \eq{cont} gives
\beq
{{ \delta \rho_Q'} \over \delta \rho_Q} = 3 (1+w_Q)
\eeq
which by the same argument as for the background is also approximately
constant at
least between about redshift one and today.
The adiabatic condition is 
\beq
{\delta \rho_Q|_{\rm adiabatic} \over \rho_Q'} = {\delta \rho_i \over \rho_i'}
\slabel{adcond}
\eeq
where $i$ is any of cold dark matter (CDM), baryons, photons or
neutrinos. Using \eq{cont} 
this can be rewritten as 
\beq
{\delta \rho_Q|_{\rm adiabatic} \over 3(1+w_Q) \rho_Q} = {\delta \rho_i \over
  3(1+w_i)\rho_i} \equiv \zeta
\slabel{adcond1}
\eeq
where $w_i$ is zero for CDM or baryons and a third for photons or
neutrinos. The final equivalence follows from the definition of
$\zeta$ and our flat gauge assumption \cite{sepuni}. Using
\eqsss{adcond1}{wQ}{Sdef}{Smeas} and that $\rho_Q \sim \rho_{\rm cdm}$ today, shows
that the adiabatic condition is strongly violated. This implies that
there must have  been more than one light degree of freedom during inflation
(see for example reference \cite{GorWanBasMaa01}). 

If we assume, as is commonly done, that the dark energy is a light
canonical scalar 
field, then its Lagrangian is given by
\beq
{\cal L_Q} = -\frac{1}{2}\partial_\mu Q \partial ^\mu Q - V_Q(Q)
\eeq
where $Q$ is the dark energy scalar field, also known as the
`quintessence', and $V_Q$ are the terms in the potential only
depending on $Q$. If during inflation, the quintessence is a light
field, then 
\beq
{\partial^2 V_Q \over \partial Q^2} \ll H_{\rm inf}^2
\slabel{light}
\eeq
where $H_{\rm inf}$ is the Hubble parameter during inflation. In which
case, $Q$ 
acquires the usual quantum fluctuations
\beq
{\cal P}_Q(k) = \left({H_{\rm inf} \over 2\pi}\right)^2
\slabel{deltaQ}
\eeq
where 
\beq
{\cal P}_x(k) \equiv {k^3 \over 2\pi^2} \left<|\delta x(k)|^2\right>
\slabel{powerspec}
\eeq
is the power spectrum of some quantity $x$ evaluated at wave number
$k$. Current observations give \cite{Sel04}
\beq
{\cal P}_\zeta^{1/2} = 5 \times 10^{-5}.
\slabel{zetameas}
\eeq
which when combined with \eqs{Sdef}{Smeas} give 
\beq
{1\over \rho_Q} {\cal P}_{\rho_Q}^{1/2} = 6\times 10^{-4}.  
\slabel{drhoQoverrhoQ}
\eeq

For a canonical scalar field, the dark energy  satisfies the
Klein-Gordon equation, which when using efolds as the time parameter is 
\beq
    Q'' + {3 \over 2} (w_T-1) Q' +{1 \over
  H^2}{\partial V_Q \over \partial Q} = 0
\slabel{KG}
\eeq
where $w_T = p_T /\rho_T$ is the total equation of state
and $\rho_T$ and $p_T$ are the total density and pressure. 
 \eq{KG} has a constant solution provided we can neglect the 
$(\partial V_Q/ \partial Q)/H^2$ term. The other solution is a decaying
mode if $w_T < 1$. The first slow parameter is defined
as 
\beq
\epsilon_Q \equiv  {M_p^2 \over 2} \left({1\over V_Q}{\partial
  V_Q\over \partial Q}
\right)^2
\slabel{epsilonQ}
\eeq 
where $M_p \equiv 1/8\pi G$ is the reduced Planck mass.
The equation of state of the quintessence is given by
\beq
w_Q \equiv { -{V_Q \over H^2} +
  {1\over 2} Q'^2 \over {V_Q \over H^2} +{1\over 2} Q'^2 }
\eeq
 which, with the  observational constraints in \eq{wQ}, implies
 \beq
 {1 \over 2}H^2 Q'^2 \ll {V_Q}
 \slabel{nokinetic}
 \eeq
 between redshift one and today.
So when $\rho_Q$
dominates the energy density
\beq
H^2 \approx {1 \over 3 M_p^2} V_Q
\slabel{friedman}
\eeq
which when combined with \eqs{KG}{epsilonQ} implies that
$\epsilon_Q<1$ is needed for an approximately constant $Q$ solution. As $H$ was
larger in the past, the conditions for constant $Q$ may be even 
better satisfied at previous times. If we assume this to be the case,
then we can neglect the third term on the left hand side of \eq{KG}. Then,
before dark energy domination, the total equation of state can be
taken to be unperturbed and using the separate Universe approach \cite{sepuni}
we
can write an approximate equation for the perturbations in $Q$
\beq
\delta Q'' + {3 \over 2 } (w_T-1)\delta Q' \approx 0.
\slabel{pertKG}
\eeq
Then, if $w_T < 1$, we only have a constant and a decaying solution. In
theory, it is still possible to have a transient period of growth in
$\delta Q$. This can happen if the initial conditions are such that
the decaying and constant mode have a different sign and the decaying
mode is initially large. However, those initial conditions are not
consistent with $Q$ being a light field whose perturbations are
generated during inflation. 

If, after inflation, there is a period in which the kinetic energy of
a canonical scalar field becomes the dominant energy source then
during this time $w_T
\approx 1$. 
Then, during the period of kinetic domination, \eq{pertKG} has a constant
and linear solution for $\delta Q$. However, between inflation and
the onset of kinetic domination, $\delta Q$ would have been constant
and so this will match onto the constant mode of the solution of
\eq{pertKG}. 

It is possible that in 
the past the quintessence had a much steeper potential and so  was not
frozen. Then its motion would have needed to carry it into a domain
were the potential slope was shallow. The `tracking' 
potentials have this property. They require \cite{SteWanZla99},
\beq
{1\over V_Q}{\partial^2 V_Q \over \partial Q^2 }\left( {1\over V_Q}{\partial V_Q \over
   \partial Q} \right)^{-2} \ge 1.
 \eeq
 They are insensitive to perturbations in the initial conditions of the
 field. This implies that any isocurvature perturbations decay
\cite{tracksup,GorHu04,Gor05}.
This makes these potentials
 unsuitable for providing
the large isocurvature perturbations needed in \eq{drhoQoverrhoQ}.


Taking the kinetic term  in the quintessence to be  negligible,
we can use $\delta \rho_Q \approx \delta Q \partial V_Q / \partial Q,$ $\rho_Q
\approx V_Q$ and \eq{epsilonQ} in  \eq{drhoQoverrhoQ} 
to get
\beq
{\cal P}_{ Q}^{1/2} = 4 \times 10^{-4} M_p \epsilon_Q^{-1/2}.
\slabel{deltaQmeas}
\eeq
As the current acceleration requires $\epsilon_Q < 1$, \eq{deltaQmeas} implies
\beq
{\cal P}_{Q}^{1/2} > 4 \times 10^{-4} M_p
\slabel{deltaQbound}
\eeq

This applies to the value of
$\delta Q$ during dark energy domination. However, as we have
discussed, a constant $\delta Q$ is a likely solution, in which case this
bound also applies to $\delta Q$ during inflation. It then follows
from 
\eqs{deltaQbound}{deltaQ} that 
\beq
V_{\rm inf}^{1/4} > 7 \times 10^{-2}M_p
\slabel{minV}
\eeq
which is inconsistent with the current observational limits on
gravitational waves \cite{Sel04}
\beq
V_{\rm inf}^{1/4} < 10^{-2} M_p 
\slabel{maxV}
\eeq
at the $95\%$ confidence level.
As proposed in \cite{MorTak03}, a way round these contradictory
bounds is to have $\delta Q$ grow between inflation and dark energy
domination. From  \eqss{maxV}{minV}{deltaQ}, we see that
we require
\beq
{\delta Q_f \over \delta Q_i} > 45
\slabel{amp}
\eeq
where subscript $i$ denotes the initial value at horizon
exit\footnote{By ``horizon exit'' we mean $a\over k$ becomes larger
  than $1\over H$ due to the growth of $a$. } during
inflation and subscript $f$ denotes the final value at dark energy
domination.


\section{Tachyonic Amplification}
\slabel{amplification}
In reference \cite{MorTak03} an amplification mechanism that relied on  varying
the coefficient of the kinetic term in the Lagrangian was
proposed. However, no specific mechanism was given 
for why the coefficient varied. Here, we propose a mechanism
that identifies the coefficient with a  scalar field which has a
tachyonic potential. Say, we write the quintessence as 
\beq
Q \equiv \phi \theta
\slabel{Qdef}
\eeq
and we assume initially $\phi$ is fixed at a constant value $\phi_i$. Then the
kinetic part of the Lagrangian can be written as
\beq
{\cal L}_{\rm kinetic} \equiv -\partial_\mu Q \partial^\mu Q = -\phi_i^2
\partial_\mu \theta \partial^\mu \theta
\eeq
and perturbations in $Q$ can be
rewritten using \eq{Qdef} as 
\beq
\delta Q_i = \phi_i \delta \theta. 
\slabel{deltaQi}
\eeq
If after inflation, and before dark energy domination, $\phi_i$ changes
to $\phi_f$ while $\delta \theta$ remains constant, then from
\eqs{Qdef}{deltaQi}, the final perturbation in the dark energy
perturbation will be
\beq
\delta Q_f = \delta Q_i {\phi_f\over \phi_i} 
\slabel{deltaQf} 
\eeq
Substituting this equation into \eq{amp} gives the bound needed to
solve the gravitational wave problem as
\beq
{\phi_f \over \phi_i} > 45.
\slabel{amp1}
\eeq
A natural way of implementing this amplification mechanism is to take
the quintessence to be a complex scalar field with a 
Lagrangian given by
\beq
{\cal L_Q} = -\partial_\mu \Phi (\partial ^\mu \Phi)^* -V(\Phi)
\eeq
where $*$ denotes complex conjugation and $V$ is the potential. Defining 
\beq
\Phi = {1\over \sqrt{2}}\phi {\rm e}^{i\theta}
\slabel{Phidef}
\eeq
then the Lagrangian is of the form
\beq
{\cal L} = -\frac{1}{2}\left( \phi^2 \partial_\mu \theta \partial^\mu \theta
+ \ \partial_\mu \phi \partial^\mu \phi
  \right) - V(\Phi).
\eeq
So that if $\theta$ is constant, $\phi$ is a canonical scalar field,
and if $\phi$ is constant then  
\beq
Q\equiv \phi \theta
\eeq
is a canonical scalar field. 

One proposed model for the quintessence is a pseudo-Nambu-Goldstone
boson \cite{FriHilSteWag95}. This has the advantage of suppressing unwanted
couplings with Standard Model fields by imposing a discrete symmetry in the
Lagrangian. In this case the radial field will have a Mexican hat type
potential
\beq
\slabel{V}
V_\phi=V_0 \left( \frac{\eta_0 \phi^2}{4M_p^2} -1\right)^2
\eeq
where $V_0$ is the potential at the maximum and $\eta_0$ is
the absolute value of the second slow roll parameter at the maximum. 
The final value of $\phi$
 is at the minimum of its potential
\beq
\phi_f = {2 M_p \over\sqrt{\eta_0}}.
\slabel{phif}
\eeq
Substituting \eq{phif} into \eq{amp1} gives the bound
\beq
\phi_i < {   M_p \over 22 \sqrt{\eta_0}}
\eeq
 needed to solve the gravitational wave problem.

\begin{figure}[tb]
\vspace{-2cm}
\centerline{
\includegraphics{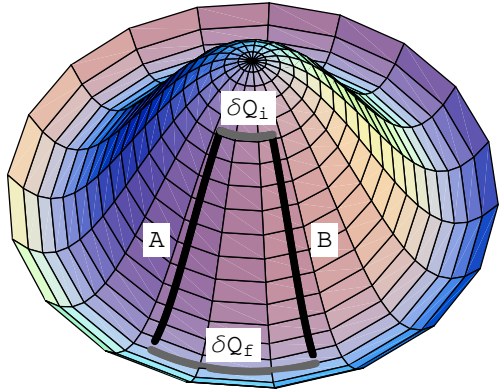}
}
\vspace{-2cm} 
\caption{\footnotesize
This figure shows how the amplification of $\delta Q$ takes place.
The Mexican hat potential, whose radial form is given in \eq{V}, is
displayed.
The large scale perturbation, $\delta Q$, can be thought of as the difference, in the circumference direction,
between two separate homogeneous Universes \cite{sepuni}, whose
trajectories  in field space (thick black lines) are labeled by `A' and `B'. The initial ($\delta Q_i$) and final ($\delta Q_f$) perturbations
are drawn  with  thick gray lines.
\label{tach}}
\end{figure}

This tachyonic amplification mechanism has also been used in
constructing models of moduli inflation \cite{KadSte03} and in the
curvaton scenario \cite{DimLytRod05} in order to get inflation at a low
energy scale.
A graphical representation of the amplification 
is given in Fig.~\ref{tach}.
\section{Radial field as the inflaton}
\slabel{onephase}
In this section we investigate whether the radial field ($\phi$), in
\eq{Phidef}, can
be used as the inflaton.
The Klein-Gordon equation for $\phi$ is
\beq
 \phi'' + {3 \over 2} (w_T-1) \phi' +{1 \over
  H^2}{\partial V_\phi \over \partial \phi} = 0.
\slabel{KGphi}
\eeq
Assuming that the potential of $\phi$ dominates the energy density,
\eq{KGphi} can
be solved by making the approximations that  
$w_T \approx -1$,
\beq
H \approx {1\over M_p}\sqrt{V_0 \over 3}
\eeq
and that $\phi$ is small enough  that we can
neglect the quartic part of the potential given by \eq{V}.
We can then obtain  the growing mode solution  
\begin{equation}
\slabel{phi}
\phi = \frac{2M_p}{\sqrt{\eta_0}} {\rm e}^{\frac{3}{2}N(1-\sqrt{1+4\eta_0/3})}.
\end{equation}
The
contribution to $\zeta$ from $\phi$ is \cite{sepuni,sepuni1}
\beq
\slabel{zetainf}
\zeta_\phi={\partial N \over \partial \phi_i } \delta\phi_i
\eeq
where subscript $i$ denotes the initial value of a quantity.
In order for $\zeta$ to be completely correlated with $\rho_Q$ we
require ${\cal P}_{\zeta_\phi} \ll{\cal P}_\zeta$.
Substituting \eq{phi} into \eq{zetainf} gives
\beq
\zeta_{\phi} = { 2\over 3(1-\sqrt{1+4\eta_0/3})}{\delta \phi_i \over \phi_i}
\slabel{zetaphii}
\eeq
The power spectrum for $\delta Q$ can be obtained by solving
\eq{zetaphii} for $\phi_i$ and substituting 
with \eq{phif} into \eq{deltaQf}. Then, as $\delta Q$ and $\delta \phi$ are
perpendicular
in field space we can use ${\cal P}_{Q_i} = {\cal P}_{\phi_i}$ in this
result to get
\beq 
{\cal P}_{Q_f}^{1/2} = M_p
{\cal P}_{\zeta_{\phi}}^{1/2}
{3(-1+\sqrt{1+4\eta_0/3})\over\sqrt{\eta_0}}. 
\eeq 
Then the bound in
\eq{deltaQbound} implies 
\beq 
{\cal P}_{\zeta_{\phi}}^{1/2} >
{4\times 10^{-4}\sqrt{\eta_0}\over 3 (-1+\sqrt{1+4\eta_0/3})}.
\slabel{zetaphibound} 
\eeq 
As $\eta_0>0$, \eq{zetaphibound} implies
\beq {\cal P}_{\zeta_{\phi}}^{1/2} > 1\times 10^{-4} \eeq 
which, as can be seen from \eq{zetameas}, is
larger than the measured total $\zeta$. It follows that the radial field is not suited to be the inflaton.

\section{Separate inflaton}
\slabel{sepinflaton}
 As seen from \eq{zetainf}, the contribution of $\phi$ to the total
$\zeta$ can be suppressed by suppressing $\delta \phi$.
 This can be achieved if $\phi$ is fixed
at a constant value while the wavelengths responsible for structure
formation are exiting the Hubble horizon during inflation. This may
come about if this era of inflation is driven by some other potential
dominated scalar field which also traps $\phi$ in an induced local
high curvature 
minimum in the potential close to the origin of $\phi$. 
At the end of this initial era of
inflation, this 
trapping mechanism would need to disappear as $\phi$ still needs, at
some point before today, to roll down to its origin so as to induce
the needed amplification in the perturbations of $Q$.

Depending on the initial value of $\phi$, it may drive a second phase
of inflation.
We have to make sure that this second
phase isn't too long or else the structure formation wave numbers will
leave the horizon during this phase and we have to face the problem
mentioned in Sec.~\ref{onephase}. 
The modes  relevant to large
scale structure have wavelengths  between one and a thousandth of the
Hubble length today. 
To estimate the number of efolds of inflation after the first of the
large scale structure modes exits the horizon we can, for most models,
neglect the change of energy scales from then until the end of
inflation. Also, assuming there
 are no phases in the Universe when $w_T >1/3$,  the number of efolds
is bounded by \cite{LidLea03}
\beq
N_* 
\lesssim 
67 - 2 \log_{10}\left({M_p \over V_{\rm end}^{1/4}
}\right) - 0.8 \log_{10} \left(\left({V_{\rm end} \over \rho_{\rm
    reheat}}\right)^{1/4} \right). 
\slabel{Nstarm1}
\eeq
Substituting the gravitational wave bound (\eq{maxV}) 
into \eq{Nstarm1} gives
\beq
N_* \lesssim 63  -  0.8 \log_{10} \left(\left({V_{\rm end} \over \rho_{\rm
    reheat}}\right)^{1/4} \right).
\slabel{Nstar}
\eeq
This upper limit is reduced by any subsequent
periods of inflation or additional periods of matter
domination. A lower bound on $N_*$ is placed by requiring that all the
structure formation modes exit during the first phase of
inflation. Combined with \eq{Nstar}, this gives
\beq
10 < N_* < 63.				   
\slabel{Nstarbound}
\eeq

Substituting \eq{phi} into \eq{amp1} and solving for N we get
\beq
N_\phi > {1 \over \eta_0} (2+\sqrt{3+4\eta_0})
\slabel{Nphi}
\eeq
where we now distinguish the efolds produced by $\phi$ with a subscript.
Any efolds of inflation due to $\phi$ will be subtracted from the
amount of efolds before the end of the first phase of inflation in
which structure formation modes left the horizon \cite{LidLea03}. This
means that in order for the structure formation modes not to exit the
horizon during any inflation provided by $\phi$, we need 
\beq
N_\phi \ll N_*
\slabel{Nrel}
\eeq
Using \eqss{Nstar}{Nphi}{Nrel} we get the limit
\beq
\eta_0 > 0.06
\eeq
which when combined with \eq{phif} implies the vev of $\phi$ satisfies
the limit
\beq
\phi_f < 8 M_p.
\slabel{phi0upper}
\eeq
The potential for $Q$ needs to be periodic and can have the form
\cite{FriHilSteWag95} 
\beq
V_Q = V_{Q,0}(1 +\cos(Q/\phi))
\slabel{VQ}
\eeq
where $V_{Q,0}$ is a constant. Then if the current phase of
acceleration is due to $Q$ slow rolling towards a minimum and we take
$\phi=\phi_f$, we need 
\beq
Q\simgt M_p
\eeq
which combined with \eq{Qdef} implies 
\beq
\phi_f \simgt M_p.
\slabel{phi0lower}
\eeq
So in order to satisfy both constraints, \eqs{phi0upper}{phi0lower} we
set  $\phi_f\sim M_p$ or $\eta_0\sim 4$. The small number of efolds of
inflation 
produced by $\phi$ and the matter dominated phase resulting from the
oscillations of $\phi$ around $M_p$ should be sufficient to dilute the
relativistic decay products of the previous era of inflation to
negligible levels. Then either $\phi$ decays to reheat the Universe
or there may be further phases of inflation at lower energy scales
such as those needed to cure the moduli problem. Any
additional inflation will subtract from the upper bound in 
\eq{Nstarbound} which means that the additional efolds should be 
less than that upper bound. Curing the moduli problem should not pose
a problem in this regard as that can be solved by about 
ten efolds of thermal inflation \cite{thermal}.

\section{Variable Decay}
\slabel{vardecay}
In the separate Universe approach, $\zeta$ can be seen to be the
perturbation in the number of efolds of expansion needed to reach a
fixed energy 
density \cite{sepuni,sepuni1}.
It follows that in order for  $\delta \rho_Q$ and
$\zeta$ to be completely 
anti-correlated we require that the amount of expansion 
undergone, to reach a fixed density,
depend on the
initial value of $Q$. In reference \cite{MorTak03} a curvaton
 based
mechanism \cite{curvaton} was proposed. This entailed adding an extra field, the
curvaton, which was taken to be strongly coupled with the quintessence
during inflation and then decoupled after inflation.

\begin{figure}[tb]
\centerline{\epsfxsize=3.4in\epsffile{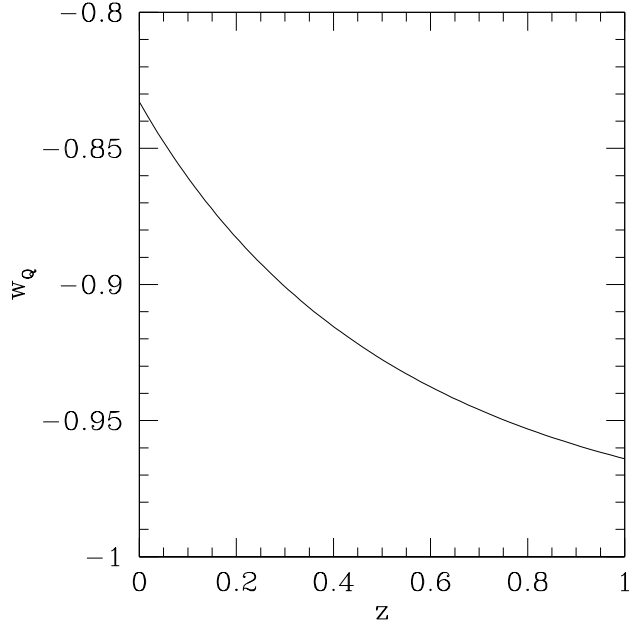}}
\caption{\footnotesize
The evolution of the  equation of state of the quintessence ($w_Q$), with 
redshift $z$, when
the potential is of the form \eq{VQ} and the initial condition is given in
\eq{Qinit}. The quintessence is taken to make up 73 \% of the energy density today. 
\label{w}}
\end{figure}

In reference \cite{GorHu04} a variable decay mechanism \cite{DvaGruzZal04,Kof03} was
used. This mechanism can be implemented by a non-renormalizable coupling between the inflaton ($\psi$), a
fermion ($q$) and
anti-fermion ($\bar{q}$) and the quintessence ($Q$):
\beq
{\cal L}_{\rm reheat} = -\left({Q\over M_p}\right) \bar{q} q \psi.
\slabel{Lreheat}
\eeq
Then, the decay rate of the inflaton into the fermions is 
\beq
\Gamma = \left({Q \over M_p} \right)^2 {m_\psi \over 8 \pi}.
\slabel{decay}
\eeq
As $Q$ is inhomogeneous, the decay rate will be inhomogeneous and at
reheating this leads to \cite{DvaGruzZal04}
\beq
\zeta = -{1 \over 3} {\delta Q \over Q}.
\slabel{zetavardecay}
\eeq
As $\delta Q / Q$ is constant, it does not matter at which point
we take this value.
Substituting \eqs{deltaQmeas}{zetameas} into \eq{zetavardecay} and
rearranging terms 
gives
\beq
\epsilon_Q \approx { 8 M_p^2 \over  Q^2}.
\slabel{epsilonQ1}
\eeq
Substituting \eqss{epsilonQ1}{VQ}{phif} into \eq{epsilonQ}, using $\eta_0=4$ and solving for $Q$ gives
\beq
Q \approx 5 M_p 
\slabel{Qinit}
\eeq
and also implies
\beq
\epsilon_Q \approx 0.3.
\eeq
The effect on the equation of state can be seen by numerically
evolving the Klein-Gordon and Friedman equations with the potential
given in \eq{VQ} and
the initial condition given in \eq{Qinit}. The result is plotted in Fig.~\ref{w}.
By comparing with the probability contours in Fig.~10 of \cite{Sel04}, it can be seen that these values are consistent with current observational bounds.

\section{Example Inflaton}
\slabel{example}
In addition to generating perturbations for large scale structure,
inflation should also
solve the flatness and 
related problems \cite{lin90}.
The simple quadratic potential in the
chaotic inflation context \cite{lin90} may be used for this
\beq
V_{\psi} = {1 \over 2} m_\psi^2
\psi^2.
\slabel{Vpsi1}
\eeq
The Klein-Gordon equation for $\psi$ can be obtained by substituting
$Q$ for $\psi$ in \eq{KG}. The slow roll approximation can be used in
this case. This entails setting $w_T=-1$, neglecting the kinetic
term contribution to $H$ and the $\psi''$ term,  giving
\beq
- \psi' + {M_p^2 \over V_\psi} {\partial \over \partial
  \psi} V_\psi =0 
\eeq
which has the solution
\beq 
\psi = M_p \sqrt{2+4 N}
\slabel{psi}
\eeq
where the efolds $N=-\log(a)$ have been set to zero when the slow
roll conditions are first violated. 
The curvature perturbation from $\psi$ can be obtained from \eqs{psi}{Vpsi1}
\beq
{\cal P}_{\zeta_\psi}^{1/2} = {\partial N \over \partial \psi} {H \over 2\pi} 
= {m_\psi \over 2\pi \sqrt{6} M_p }(1+2N ).
\eeq
In order to have the perturbations from $\psi$ sub-dominant we require
${\cal P}_{\zeta_\psi} \ll {\cal P}_\zeta $ which using
\eqs{zetameas}{Nstarbound}  implies
\beq
m_\psi \simlt 10^{-5}M_p.
\slabel{mpsi}
\eeq
During this time, the $\phi$ field needs to be trapped near its
origin. This can be achieved by
adding a coupling term to the
potential in \eq{Vpsi1} \cite{KofLin86}
\beq
V = {1 \over 2} \left(m_\psi^2 +g^2(\phi -\phi_i)^2 \right)\psi^2 
\slabel{Vmod}
\eeq 
where $\phi_i$ is the trapping point and $g$ is a dimensionless coupling constant.
The trapping also requires that the $\phi$ be heavy:
\beq
 {\partial^2 V \over \partial \phi^2}  \gtrsim  H^2
\eeq 
which, when used with \eq{Vmod} and $\phi=\phi_i$, implies
\beq
m_\psi \lesssim \sqrt{6} g M_p.
\slabel{mpsibound2}
\eeq
In order for the coupling not to change the inflaton potential through radiative corrections, $g<10^{-3}$ is required \cite{lin90,lin92}. It follows that \eq{mpsibound2}
is easily satisfied due to
\eq{mpsi}.

During this first phase of inflation, the quintessence is 
virtually a free field and it acquires the usual quantum fluctuations,
which using \eqs{psi}{Vpsi1} are given by
\beq
{\cal P}_{Q_i}^{1/2} = {H_*\over 2\pi} ={ m_\psi \sqrt{1+2N_*} \over 2\pi
  \sqrt{3} }
\slabel{deltaQ0}
\eeq 
where subscript $*$ denotes the value of a quantity at horizon crossing.

If $Q$ is the variable decay field, it follows from \eq{Lreheat} that
in order for quantum corrections not to significantly modify \eq{Vpsi1} 
we need \cite{lin90}
\beq
Q_i^2 \lesssim 10^{-5} M_p^2.
\slabel{Q0bound}
\eeq
It is important to check that this is not smaller than 
 the perturbations in \eq{deltaQ0}, else the perturbation
will be suppressed \cite{lin84}. Using 
\eqs{deltaQ0}{Nstarbound} gives
\beq
{{\cal P}_{Q_i}^{1/2} \over Q_i} < {m_\phi \over Q_i}
\eeq
which shows $\delta Q \ll Q$ is compatible with the bounds in
\eqs{Q0bound}{mpsi}.  
Also, if we take $\phi_i \sim Q_i$ then from \eq{Q0bound} and \eq{phi}, with
$\eta_0 = 4$, the amount of inflation from $\phi$ is bounded by
\beq
N_\phi \gtrsim 3.
\slabel{Nphia}
\eeq
The density at reheating  is given by solving 
the decay rate given in \eq{decay} to be
 $\Gamma = 3H_{\rm
  reheat}$ with the constraints \eqs{Q0bound}{mpsi} to give 
\beq
\rho_{\rm reheat}^{1/4} < 10^{-6} M_p. 
\eeq
Substituting this into \eq{Nstarm1} and using \eqs{Vpsi1}{mpsi} reduces the
upper bound of $N_*$ by about one. Combined with the reduction due to the
inflation caused by $\phi$, \eq{Nphia}, this gives
\beq
10 < N_* < 58.				   
\slabel{Nstarbound1}
\eeq

Finally, we can estimate the scalar spectral index. We
can set $\eta_0$ to zero as we are assuming the radial field ($\phi$)
is trapped at $\phi_i$ while the structure formation modes are
exiting the horizon. The spectral index is given by
\beq
n_s-1 \equiv  {\partial  \log({\cal P_\zeta}) \over \partial \log(k) }
\slabel{ns}
\eeq
 As we are assuming the perturbations from $\psi$ contribute
 negligibly to $\zeta$, it follows that
 $\zeta$ is completely spatially anti-correlated with
$\delta Q$ and so will have the same spectral index as $\delta Q$.
As ${\cal P}_Q$ depends on the value of the Hubble parameter, \eq{deltaQ},
     when $k = {\rm e}^{-N} H$, and $H'\ll
     H$,
 we can rewrite \eq{ns}  as
\beq
n_s-1 ={ 2\partial \log(H) \over \partial 
  \log({\rm e}^{-N} H)} \approx -2{ \partial \log(H) \over \partial 
  N}.
\slabel{ns2}
\eeq
Then, substituting \eqs{Vpsi1}{psi} into \eq{ns2} gives
\beq
n_s = 1 - {2 \over 1+2N}.
\slabel{ns3}
\eeq
The current observational bound on the spectral index is \cite{Sel04}
\beq
n_s > 0.94 \quad (97.5\% \mbox{ C.I.})
\eeq
which when combined with \eq{ns3} 
implies that 
\beq
N_* >16
\slabel{Nlowerbound}
\eeq
and using the upper bound in \eq{Nstarbound1} in \eq{ns3}
\beq
n_s < 0.98
\eeq
to be consistent with the inflaton potential in \eq{Vpsi1} in this
tachyonic amplification scenario.

\section{Conclusions}
\slabel{Conclusions}
In this article we have explained a mechanism for amplifying dark
energy perturbations. The motivation for this was that sufficiently
large dark energy perturbations which are anti-correlated with the
other perturbations provide a possible explanation for the low CMB TT
quadrupole. However, if the perturbations are generated during
inflation and frozen until today then they are not of sufficient size. 

We proposed using a tachyonic amplification mechanism
to overcome this problem. The dark energy
is modeled as a complex scalar field whose radial component has a
Mexican hat potential. The part of the dark energy (the
`quintessence') that causes the 
acceleration today is modeled as a pseudo-Nambu-Goldstone boson.

It was shown that there is a difficulty in using the radial component
of the dark energy as the inflaton. This is because it then produces a
non-negligible curvature perturbation which is uncorrelated with the
quintessence perturbation. We show that a working mechanism can be
achieved  if
the inflaton is a separate field and the radial component is trapped
near the maximum of its potential during inflation. After inflation,
the trapping mechanism needs to be released so that the radial
component can roll down to its minimum. This change in the radial
component provides the necessary amplification.

We also showed how the anti-correlation can be achieved by using the
variable decay mechanism. By coupling the quintessence to the
inflaton and a fermion/anti-fermion pair, the perturbations in the
quintessence field cause perturbations in the curvature perturbation
by varying the time at which the inflaton decays. We showed how this
then constrains the quintessence potential and  initial
value of the quintessence.

Finally, we looked at the quadratic potential as an example
inflaton. We showed how it could satisfy all the necessary constraints
and we put limits on the spectral slope and the duration
of the initial period of inflation.

The current statistical significance of the detection of dark energy
 perturbations 
 is modest, however, future CMB polarization data may improve this
 \cite{GorHu04}. It may also be possible to use CMB polarization and
 polarization/temperature cross-correlation data to distinguish
 between dark energy perturbations and other explanations for the low
CMB  temperature quadrupole \cite{GorHu04}.
 A detection of a non-homogenous dark energy
 would provide a valuable new
window on both the nature 
of the dark energy and the process which generates the primordial
perturbations. 

\smallskip{\it Acknowledgments:}
We thank D.~Chung, G.~Dvali, W.~Hu, D.~Lyth and J.~Frieman for helpful
discussions. CG  is supported by the KICP under NSF PHY-0114422.

\end{document}